\documentstyle[emulateapj,onecolfloat,psfig]{article}

\newlength{\tskip}
\newlength{\colwidth}

\newcommand{\beq}{\begin{equation}}
\newcommand{\eeq}{\end{equation}}
\newcommand{\beqa}{\begin{eqnarray}}
\newcommand{\eeqa}{\end{eqnarray}}
\newcommand{\lsim}{\lesssim}
\newcommand{\gsim}{\gtrsim}
\newcommand{\psim}{\mbox{\raisebox{-1.0ex}{$~\stackrel{\textstyle \propto}
{\textstyle \sim}~$ }}}

\newcommand{\lmk}{\left(}
\newcommand{\rmk}{\right)}

\begin{document}
\twocolumn[
\title{Can LISA Resolve Distance to the Large Magellanic Cloud?}
\author{Asantha Cooray$^1$ and Naoki Seto$^2$}
\affil{$^1$Department of Physics and Astronomy, University of California, Irvine, CA 92617\\
$^2$Theoretical Astrophysics, Division of Physics, Mathematics and Astronomy, 
California Institute of Technology, MS 130-33, Pasadena, CA 91125\\}

\begin{abstract}
The Laser Interferometer Space Antenna (LISA) is expected to detect $N \sim 22\times 10^{\pm1}$ 
close white dwarf binaries in the Large Magellanic Cloud  (LMC)
through their gravitational radiation with signal-to-noise ratios greater than $\sim$10 in observational durations of
3 years or more. In addition to chirp mass, location on the sky, and
 other binary 
parameters, the distance to each binary is an independent parameter that
 can be 
extracted from an analysis of gravitational waves from these binaries. Using a sample of binaries, one
can establish the mean distance to the LMC as well as the variance of this distance.
Assuming no confusion noise at frequencies above 2 mHz,
LISA might determine the LMC distance to $\sim$ 4.5$\sqrt{N/22}$\% and the
line of sight extent of LMC to $\sim$ 15$(N/22)^{1/4}$\%, relative to its distance, at the one-sigma
confidence. These estimates are competitive to some of the proposed direct geometric techniques to measure
LMC distance in future with missions such as SIM and GAIA.
\end{abstract}

]

\section{Introduction}
The distance to the Large Magellanic Cloud (LMC) is the first-step in the extragalactic distance scale.
Its distance has been estimated using over 80 
different techniques, among which red clump stars, Cepheids, RR Lyrae stars,
cluster main sequence fitting, and eclipsing binaries are the well-known (see, a summary in
Benedict et al. 2002). While a combined average of recent results from selected ``best'' techniques  
indicate a distance modulus of 18.50 $\pm$ 0.02 (Alves 2003) or a few percent accurate distance to
LMC, no single technique
has reached the precision to be a reliable determination on its own. When considered as a whole, various
 techniques suggest the existence of two distinct distance scales:
a ``short-scale'' with a modulus at the low end below 18.5, and a ``long-scale'' above this 
mean (Jensen et al. 2003).

The HST Key Project calibrated the Cepheids to a LMC 
distance modulus of 18.5 with a systematic uncertainty of 0.13 mag (Freedman et al. 2001). However, it is suggested 
that the measurement scatter 
would be larger than 0.13 mag even for best results from techniques that give a modulus close to 18.5
(Jensen et al. 2003). While the uncertainty to the LMC distance can be
reduced with better data and with elimination of 
systematic effects, it is still useful to consider
new methods that have the potential to make significant improvements in the future and are
less affected by systematics and uncertain calibrations. 

The biggest improvement is expected for methods that can produce a direct geometric distance measurement to LMC
without any need to cross-calibrate against techniques that depend on the distance ladder. 
Several examples are discussed in Gould (2000)
and involve the trigonometric parallax and estimates based on kinematic arguments. 
The LMC parallax, at the level of 20 $\mu$as, can be established
with the Space Interferometer Mission (SIM) to 
an accuracy of 2 $\mu$as to 8 $\mu$as depending on the number of repeated observations.
This leads to, at best, a 10\% distance determination to the LMC.
Among the kinematic method, the geometric distance based on the 
light travel time across the SN 1987A ring is now well known
(Panagia et al. 1991), but differences 
at the 10\% level in distance still remain on the interpretation of the same data
(Gould 1995; Sonneborne et al. 1996). The second kinematic method based on the radial velocity gradient technique (e.g., Detweiler et al. 1984) requires measurements of the LMC radial velocity field, its mean proper motion, and the position angle of LMC photometric nodes (Gould 2000). 
The limitation remains with the LMC proper motion measurement. While an estimate accurate to 2\% was
expected with the Full-Sky Astrometric Mapping Explorer (FAME),  due to the cancellation of this project by NASA,
the next opportunity relies on the launch of
GAIA\footnote{http://astro.estec.esa.nl/GAIA}.  

Around the same time scale as results from GAIA would be available,
there is another technique to establish the distance to LMC directly. This technique involves the analysis of
gravitational waves from a sample of close white dwarf binaries  (CWDBs)
in LMC using observations with the 
 Laser Interferometer Space Antenna
(LISA)\footnote{http://lisa.jpl.nasa.gov} mission. 
As studied in the literature (Hils et al. 1990;
Nelemans et al. 2001), the LISA's frequency coverage between 10$^{-1}$
Hz and 10$^{-4}$ Hz is ideal 
for a direct detection of CWDBs in our galaxy, in addition to a large
number of possible extragalactic binary sources such as merging  massive
black holes   (Hughes 2002).  While a variety of Galactic
binaries could 
be expected, including those involving 
neutron stars,  solar-mass scale black holes or some interacting
binaries, we will  
only consider the sample related to binary white dwarfs as these are
expected to be the major fraction of GW emitting binaries with
relatively large amplitude and simple evolution. 

At frequencies above $\sim 2$ mHz, most CWDBs will be spectrally resolved (Cornish \& Larson 2003), 
and one expects a detection of $\sim 2200\times10^{\pm1} (f/2 \; {\rm
mHz})^{-8/3}$ 
galactic binaries (e.g., Hils et al. 1990; Bender \& Hils 1997; Seto
2002 and references 
therein) from the disk alone.  Note that there is an order of magnitude
uncertainty in the expected number of binaries in either  direction (Bender \& Hils 1997).  
%Cooray et al. (2003) discusses a possibility related to optical identification of these binaries 
%via follow-up observations of LISA error boxes.
Scaling to a LMC mass of $(8.7 \pm 4.3) \times 10^9$ M$_{\sun}$ (van der Marel et al. 2002), 
assuming a mass ratio of 1\% relative to the Milky Way ($\sim 5 \times 10^{11}$ M$_{\sun}$; Kochanek 1996), 
we expect $\sim$22 CWDBs in the LMC to be resolved by LISA above 2 mHz.
This number is again uncertain by an order of magnitude and the fraction relative to Milky Way is
further complicated by the fact that the star-formation histories of the two disks 
are different.
While the current LMC star-formation rate is a factor of 2 to 5 higher than the Milky Way, it is unlikely that this
will substantially increase the number of CWDBs in LMC relative to the Galaxy as the binary formation traces the past
history with a time-lag of order a few Gyrs or more. 
%The different is star-formation history has been used to explain
%the relative increase in high-mass X-ray binaries in the LMC relative to the galaxy 
%(Shtykovskiy \& Gilfanov 2005).

The gravitational waves from each binary allow one to establish certain parameters related to that binary, such
as the location, chirp mass, distance, period, and the orientation or the direction of the binary 
with respect to LISA.  Here, we focus on the radial distance measurement, but our analysis considers
measurement of all these parameters from the data streams as these
parameters are correlated with  each other. In the case of the Galactic structure,  
Ioka et. al. (1998) performed an analysis similar to the current approach.

The discussion is organized as following: In the next section, we 
consider the distance measurement with LMC CWDBs in the LISA data stream
and how it can be applied in the context of general LMC studies.
While with LISA distance information can be directly extracted from the
gravitational-wave signal,   the location information on the sky one can
obtain is limited due to relatively small signal-to-noise ratios
(SNRs).   
This could result in a substantial uncertainty in locating each 
binary within the LMC. It would, however, still be easily possible 
to distinguish LMC binaries from
Galactic ones through information on the estimated distances, as
the Galactic scale $\sim 10$kpc is much smaller than the distance to LMC
$\sim 50$kpc.  In addition to the mean 
distance estimated from
identified  CWDBs in LMC, one can also determine the radial
line-of-sight extent of LMC. 
The line-of-sight distance through LMC is
needed when interpreting microlensing observations (e.g., Mancini et al. 2004)
and previous modeling shows the existence of a number of Cepheids both behind and
front of the main disk at distances $>$ 7 kpc (Nikolaev et al. 2004).
Given that one expects LMC to be tidally disrupted, three-dimensional
information is also useful when modeling the origin and nature of Magellanic clouds in 
general (e.g., Bekki \& Chiba 2005). We conclude with a summary in \S~3.

\begin{figure*}[t]
\centerline{
\psfig{file=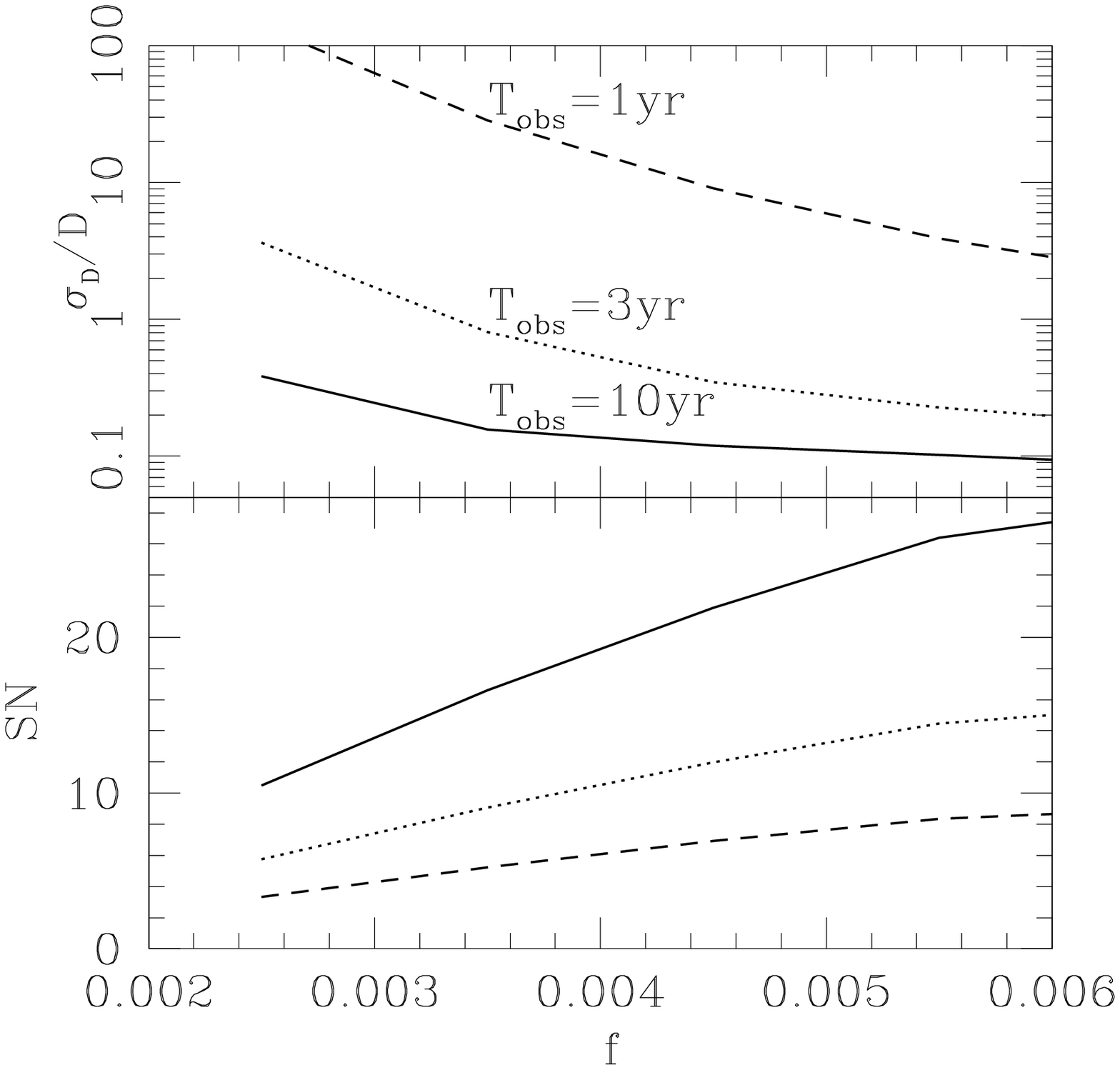,width=3.6in,angle=0}
\psfig{file=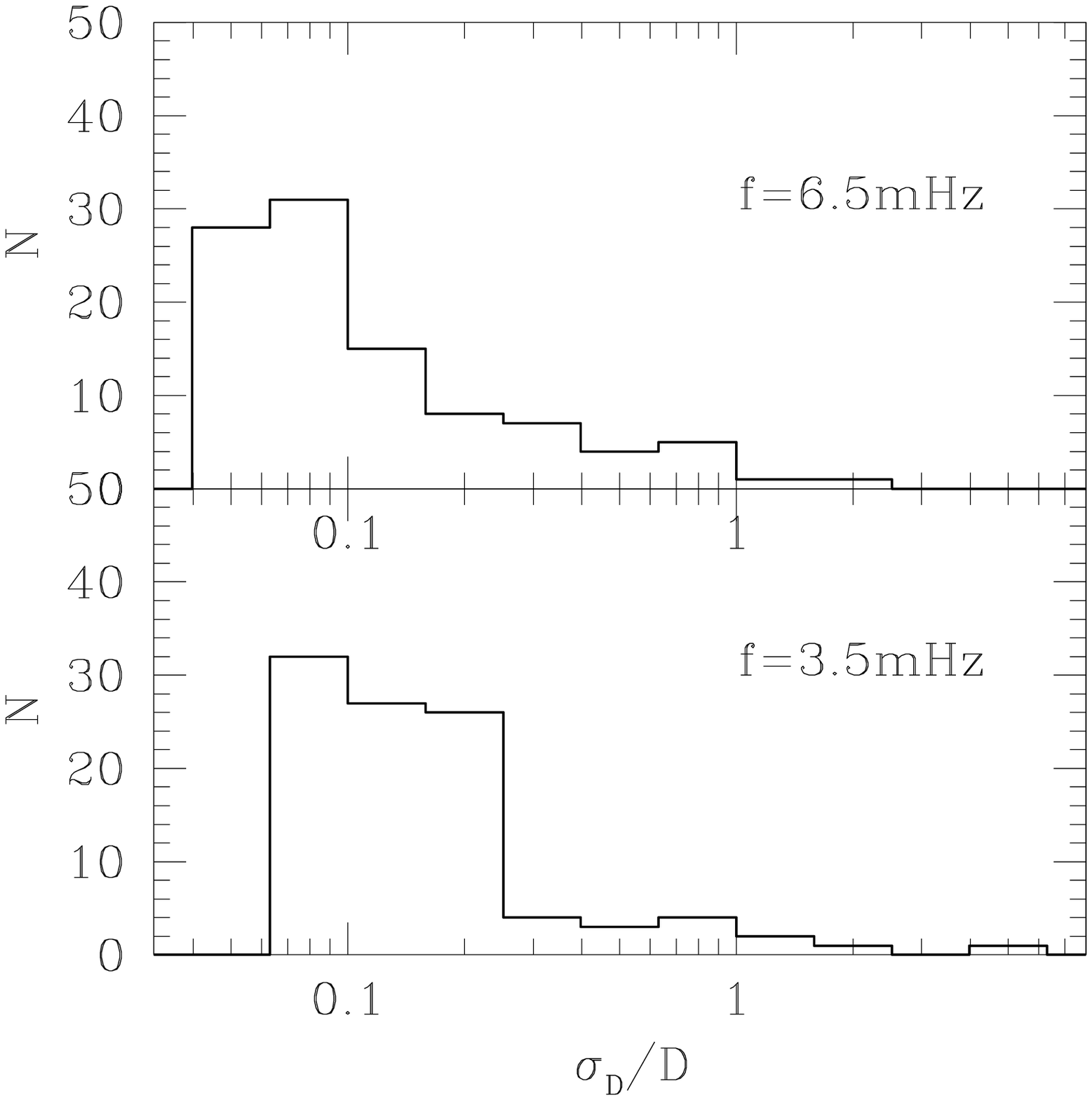,width=3.6in,angle=0}
}
\caption{ Distance determination with LISA. Here, we perform a Monte Carlo
calculation at 5 gravitational wave frequencies between 2.5 mHz and 6.5 mHz
with 100 CWDBs at each of these frequencies. We assume a  mean distance to LMC of 50 kpc.
{\it Left panel:} We show the median error in the distance measurement for this sample, 
relative to the LMC mean distance, as a function of the gravitational wave frequency. From top to bottom, the
 three curves show the mean for LISA observations over one, three, and ten years, respectively.
On average, with observations that span 10 years, we can determine distance 
 to an accuracy of 10\%, though for short-term observations, the distance determination becomes impossible.
Observations that span at least 3 years are required.  For comparison,
we also show the signal-to-noise ratio which simply scales as $SNR\propto T_{obs}^{-1/2}$.   {\it Right panel:} We show the 
distribution of fractional distance errors for the whole sample of 100 CWDBs observed
with LISA over 10 years at two representative frequencies of 3.5 mHz and 6.5 mHz.
The distribution of errors is highly skewed with a tail than span to estimates that are
uncertain at the level of 100\% or more. This tail reflects the inclination angle
of the binary sample with respect to LISA; distance determination to binaries
that are face-on becomes highly uncertain.
}
\label{fig:area}
\end{figure*}

\section{LISA Studies of LMC}

First, we will review the CWDB detections with LISA and the concentrate on the
distance measurement related to LMC. 

A CWDB is expected to have a circular orbit due to tidal interaction in
its early evolutional stage.  
The characteristic  amplitude of its gravitational waves is given by (Thorne 1987)
\beqa
A &=& 2  \frac{G^{5/3}}{D c^4} (\pi 
f)^{2/3} M_c^{5/3} \\ 
&=&1.5\times 10^{-24} \lmk\frac{M_c}{0.3M_\odot}  \rmk^{5/3}
\lmk\frac{50 {\rm kpc}}{D}  \rmk   \lmk\frac{f}{\rm 10^{-3} Hz}  \rmk^{2/3}     \, , \nonumber
\label{amplitude}
\eeqa
where $D$ is the distance to the source, and $M_c$ is the chirp mass
defined by $M_c=(M_1 M_2)^{3/5}(M_1+M_2)^{-1/5}$ with two masses $M_1$
and $M_2$ of the binary.  We can also estimate the chirp mass through
observation of the  time evolution of the frequency $\dot f$ as it is
given by 
\beqa
{\dot f}&=&\frac{96\pi^{8/3} G^{5/3}}{5c^5} f^{11/3}M_c^{5/3} \\
&=& 7.9\times 10^{-19} \lmk \frac{f}{\rm 10^{-3}Hz} \rmk^{11/3}
\lmk \frac{M_c}{0.3 M_\odot} \rmk^{5/3} {\rm sec^{-2}} \, . \nonumber
\eeqa
The resolution $\Delta {\dot f}$ depends strongly on the observational
period $T_{obs}$ as $\Delta {\dot f}\psim T_{obs}^{-5/2}$.
From eqs.(1) and (3) we can estimate the distance $D$ from the measured
amplitude $A$, frequency $f$, and its time derivative $\dot f$ as
\beq
D=\frac{5c {\dot f}}{48\pi^2 A f^3}.
\eeq
This important fact related to the distance measurement was pointed out by Schutz in 1986 (Schutz 1986).
 In the case of an adequate sample of GW emitting binaries in the LMC,  one can determine both the
mean distance as well as the width.

To evaluate the estimation error for distance of  each binary with LISA,
we have  to determine at least 8 parameters 
 (including, $A$, $f$ and $\dot f$) concurrently from the LISA data
stream that is affected by the complicated motions of three LISA satellites. 
 These parameters are the direction of the binary (two 
parameters), orientation of the binary   (two
parameters) and a phase constant. 
We refer the reader to the literature (Cutler 1998;
Takahashi \& Seto 2002; Rubbo et al. 2004; Kr{\' o}lak et al. 2004; Vecchio \& Wickham 2004)
for detailed studies related to  parameter estimation of nearly monochromatic binaries.

In this paper we used a code based on Seto (2004) that was originally
written  for super massive black hole binaries. In this code, the
parameter estimation errors are evaluated with the Fisher matrix
approach for LISA's three orthogonal data streams used  for a data analysis technique
 based on the time delay interferometry (TDI; Armstrong et al. 1999; Prince et
al. 2002). This code includes the complicated effects caused by the
finiteness of the arm-length of the detectors (see, Seto 2002;
Cornish \& 
Rubbo 2003; Rubbo et al. 2004; Vecchio \& Wickham 2004).  For the instrumental noise curve 
of LISA, we use the standard values given in Prince et al. (2002) and do
not include the binary Galactic  confusion noise as we are dealing with
binaries  at $f\gsim 2$mHz. In the last part of this section, we will return to the issue
of confusion noise and will discuss its role in changing our basic results.

To obtain how well LISA can establish distance from a sample
of CWDBs at the distance to LMC, taken to be 50 kpc,
we performed Monte-Carlo analysis at five frequencies between
2.5 mHz and 6.5 mHz with each frequency containing a sample of 100
binaries with  fixed  chirp mass  at
0.3 M$_{\sun}$. If we normalize the total number of binaries to be 22, we
expect 15 binaries at 2.5mHz, 4 at 3.5mHz, 2 at 4.5mHz, and 1 at 5.5mHz. 

In Fig.~1, we summarize our results with respect to the distance
measurement and signal detection. With a one year integration, we have a
typical $SNR$ value of $\lsim 10$ and it would not be easy to detect CWDBs in the LMC. 
Furthermore, the
distance estimation error is  considerably large. In the top panel of the left
figure,  
the  distance error is given in a relative form $\sigma_D/D$ with respect to the mean distance $D$.
For one year data, this ratio is larger than 1 and no constraints on the distance possible. 
This $1$ year result should be regarded as a reference given that the
 Fisher matrix approach we use here is  based on the linear response of the fitting parameters to the data. 
The error  $\sigma_D/D$  becomes significantly smaller with
$T_{obs}=3$yr. This is partly due to the decrease in the correlation between parameters when $T_{obs}\gsim 2$yr, and
partly due to the rapid improvement on the estimation of $\dot f$ as a
function of $T_{obs}$  (Takahashi \& Seto 2002).

In the right panel of figure 1,  the histogram of the distribution of
the distance estimation errors are given for 100 binaries with
$T_{obs}=10$yr at two specific frequencies. We can observe a tail at
large $\sigma_D/D$. It is made by nearly face-on binaries. In this
highly symmetric configuration,  it is difficult to determine the two
parameters for the
direction of the angular momentum accurately. These parameters  have a
large 
correlation with the GW amplitude $A$ under the Fisher matrix formalism  
(Takahashi \& Seto 2002),  though the SNRs  for binaries at a given
location become  maximum at  
face-on configuration. 

For each CWDB in the LMC, we determine the
distance error $\sigma_{D,i}$. 
Under the hypothesis that all binaries are at the same distance,
we estimate the variance on this distance as $\sigma^{-2}_{\bar{D}}=\sum 1/\sigma_{D,i}^2$.
To quote final errors, we renormalize the whole binary sample in our Monte Carlo calculation
 to an average number of LMC CWDBs of 22. Following this procedure, on average, we find 
that the mean distance can be established to 4.5$\sqrt{N/22}$\% at the one-sigma confidence level.
The measurement is competitive with some of the suggested techniques to establish the distance scale to LMC
based on a single technique and is likely to be at the same level as the one based on the radial velocity
gradient technique using GAIA's proper motion estimate. Incidently, in addition to the mean distance,
we can also determine the line-of-sight width across LMC based on our sample of CWDBs.
This involves estimating the excess variance of the distance estimates. To simplify the procedure,
we ignore complications resulting from the true three-dimensional structure of LMC and make use of the hypothesis
that the width is zero. We estimate the
standard deviation on the distance errors  and quote this as the uncertainty to which the thickness,
$\Delta D$,  can be measured. The error is calculated following $\sigma^{-4}_{\Delta D}=\sum 2/\sigma_{D,i}^4$.
For the same sample of 22 CWDBs, we find the error on the width to be at the level of 15$(N/22)^{0.25}$\% at the one-sigma
confidence level relative to the mean distance of LMC. With the mean distance at 50 kpc, this amounts
to establishing the thickness at 7.5 kpc.  

While the mean distance is useful to establish the cosmic distance scale, the thickness addresses
 an important cosmological problem. While several microlensing surveys have monitored LMC, 
the observational data, with a measured optical  depth of 12$^{+4}_{-3}
\times 10^{-8}$ (Alcock et al. 2000),  is inconsistent with various model expectations (see, for example, Sahu
2003) which 
indicate a lower optical depth. The differences can be reconciled if there is a significant
stellar populations either in the background or foreground of LMC such that  self-lensing,
lensing of LMC stars by sources within LMC, become important. While there is no convincing
evidence for such massive structures based on the data so far (see, van der Marel 2004), this
possibility is still not ruled out.  The three dimensional analysis of LMC by
Nikolaev et al. (2004) suggest the presence of
Cepheids both behind and in front of the main LMC disk at distances excess of 7 kpc. The width that can be
determined from CWDBs in the LISA data is comparable to such a line of sight extent. Thus, LISA data
may play a crucial role in further understanding the three-dimensional structure of LMC, especially
if the sample of CWDBs detectable with LISA were to be higher than our estimate.

In addition to LMC, LISA data can also be used to constrain the structure of SMC. With a mass of
$3 \times 10^9$ M$_{\sun}$ (Gardiner \& Noguchi 1996), we renormalize to a binary fraction that is
a third of LMC. Scaling our numbers to SMC distance of $\sim 60$ kpc with
$\sim$ 7 binaries, we find that the mean distance can be established to $\sim$ 9\% and the
line of sight extent of SMC to the level of $\sim$ 30 kpc, both at the one-sigma level. These constraints
may allow one to study the complex SMC line of sight structures where Cepheids have been observed
across a line-of-sight depth of $\sim$ 30 kpc (Mathewson et al. 1986).

So  far we have not included the effects of  the astrophysical confusion
noise. The 
binaries that mainly contribute to determine the mean distance or the
width  is relatively at high frequencies with $f\gsim
4.5$mHz.  Therefore, our result would not be significantly
affected by 
the Galactic white-dwarf binary confusion noise (Nelemans et al. 2001).
However, we note that the effective noise level at these high frequencies could 
increase by a factor $\sim 1.5$ due to  the fitting residual of resolved
binaries for certain model parameters (see {\it e.g.} figure 6 in Barack
\& Cutler 2004).  

There is another source of confusion 
noise that is worth mentioning here.  This is the cosmological GW background made by
inspiral waves  from  extreme mass ratio binaries with systems made  by  compact
objects (with mass $\sim 1 M_\odot$) and a  supermassive black holes.  The gravitational wave
signals from such binaries are highly complicated due to the effects of strong gravity though the amplitudes are
weak as these binaries are found at cosmological distances. 
Barack \& Cutler (2004) pointed out that such binaries, however, may contribute to an additional
confusion noise and could increase  the effective noise level of LISA, even in the
high frequency range around $5$ mHz. 
At present,  there are large uncertainties in calculating the amplitude
of the background made by the  extreme
mass-ratio binaries. If this 
background is fairly large, the detection  of LMC binaries themselves could be impacted  given that
these binaries are detected with marginal SNRs ($\sim 10$) in the presence of no confusion.
%It is also likely that an improved data analysis procedure would be developed to remove such a background.
%While we do not include this background, we caution the reader that this remains the last uncertainty 
%in our estimates
%related to how accurately LISA can resolve the distance to LMC.

\section{Summary}

The Laser Interferometer Space Antenna (LISA) is expected to detect a few to few hundreds
close white dwarf binaries in the Large Magellanic Cloud  (LMC)
through their gravitational radiation. The distance to LMC is an
independent parameter that can be 
extracted from an analysis of gravitational waves from these binaries.
Taking a reasonable estimate on the number of CWDBs that can be resolved with LISA
above  a gravitational wave frequency of 2 mHz to be $\sim$ 20, we find that  
LISA might determine the LMC mean distance to $\sim$ 4.5\% and the
line of sight extent of LMC to the level of 7.0 kpc, both at the one-sigma level. 

\vspace{0.5cm}

{\it Acknowledgments:} 
This work has been supported at Caltech by NASA grant NNG04GK98G and the Japan Society for the
Promotion of Science (NS). We thank Mike Kesden for helpful discussions related to statistics.

\end{document}